\documentstyle[11pt,fullpage]{article}
\renewcommand{\Pr}[0]{\mbox{\rm Prob}}
\renewcommand{\Pr}[1]{\mbox{\rm Prob}\left( #1 \right)}
\renewcommand{\Pr}[1]{\mbox{\rm P}\left( #1 \right)}

\newcounter{protoline}
\newlength{\boxwidth}

\newtheorem{proto_body}{Protocol}[section]
\newenvironment{protocol}[1]{ 
\setcounter{protoline}{0}
\begin{minipage}{\boxwidth}

\begin{proto_body}[ #1 ] \hspace*{\fill}\par
\begin{description} }{\end{description}\end{proto_body}\end{minipage}}

\newenvironment{protocol_cont}[0]{ 
\begin{minipage}{\boxwidth}
\addtocounter{proto_body}{-1}

\begin{proto_body}
\ \\ 
\begin{description} }{\end{description}\end{proto_body}\end{minipage}}

\newcommand{\indtheorem}{\begin{quote}\begin{theorem}}
\newcommand{\indtheoremend}{\end{theorem}\end{quote}}
\newcommand{\indlemma}{\begin{quote}\begin{lemma}}
\newcommand{\indlemmaend}{\end{lemma}\end{quote}}

\newcommand{\bra}[1]{\langle #1 |}
\newcommand{\ket}[1]{| #1 \rangle}

\newcommand{\comment}[1]{}

\newcommand{\trp}[2]{\mbox{Tr}_{#1}(#2)}

\newcommand{\Sa}{\mbox{${\cal A}${\sl lice\/}}}
\newcommand{\Saa}{\mbox{${\cal A}${\sl lbert\/}}}
\newcommand{\Ra}{\mbox{${\cal B}${\sl ob\/}}}

\newcommand{\pp}[2]{\mbox{pp.~#1\,--\,#2}}

\newcommand{\jofc}{{\it Journal of Cryptology}}

\newcommand{\crypto}[1]{{\it Advan\-ces in Cryptology ---
Proceedings of Crypto\,'#1}, \aou~19#1, Springer\,--\,Verlag}
\newcommand{\euroc}[2]{{\it Advanc\-es in Cryptology ---
Proceedings of Eurocrypt\,'#1}, #2~19#1, Springer\,--\,Verlag}
\newcommand{\focs}[2]{{\it Proceed\-ings of #1 \mbox{Annual} IEEE Symposium
on the Foundations of Computer Science}, #2}

\newcommand{\aou}{\mbox{August}}

\newcommand{\nov}{\mbox{November}}
\newcommand{\dec}{\mbox{December}}

\begin{document}
\hyphenation{pre-sents pre-sent}

\title{A brief review on the impossibility of quantum bit commitment}

\author{
Gilles Brassard\ \ \ \ \ \ \ Claude Cr\'epeau\,\\
Universit\'e de Montr\'eal
\thanks{\,D\'epartement IRO, Universit\'e de Montr\'eal,
C.P. 6128, succursale centre-ville,\hfill\mbox{}
\mbox{Montr\'eal (Qu\'ebec), Canada H3C 3J7.  e-mail:
\{brassard,crepeau\}@iro.umontreal.ca.}}
\and
Dominic Mayers\,\\
Princeton University
\thanks{Department of Computer Science, Princeton University, \hfill\mbox{} 
\mbox{Princeton, NJ 08544-2087, U.S.A..   e-mail:
mayers@cs.princeton.edu.}}
\and
Louis Salvail\,\\
BRICS
\thanks{BRICS, Basic Research in Computer Science of
	the Danish National Research Foundation,
      Department of Computer Science,University of \AA rhus,
Ny Munkegade, building 540, 
DK-8000 \AA rhus C, Denmark.\hfill\mbox{}\mbox{e-mail: salvail@daimi.aau.dk}}
}

\date{December 10, 1997}

\maketitle

\abstract{ 
The desire to obtain an unconditionally secure bit commitment protocol
in quantum cryptography was expressed for the first time thirteen
years ago.  Bit commitment is sufficient in quantum cryptography to
realize a variety of applications with unconditional security.  In
1993, a quantum bit commitment protocol was proposed together with a
security proof.  However, a basic flaw in the protocol was discovered
by Mayers in 1995 and subsequently by Lo and Chau.  Later the result
was generalized by Mayers who showed that unconditionally secure bit
commitment is impossible.  A brief review on quantum bit commitment
which focuses on the general impossibility theorem and on recent
attempts to bypass this result is provided.  }

\section{Introduction} After that Mayers obtained his general impossibility
theorem for bit commitment (see the Appendix and
\cite{dom:impos,mayers97}), different kind of ideas were proposed by
Brassard, Cr\'epeau and Salvail with the hope to realise
unconditionally secure bit commitment \cite{bc96}.  It was then
realized by Mayers that these apparently promising ideas were also
ruled out by his attack.  These attempts contributed to enhance our
understanding of what is going on with quantum bit
commitment\cite{claude}.  However, no complete discussion on the
subject has ever been provided in the litterature.

Furthermore, two different proofs, each using a different approach,
was provided by Mayers.  The first approach was used in the original
proof (see the Appendix and \cite{dom:impos}) whereas the second
approach appeared later in \cite{mayers97}.  
Despite all these results, and the related discussion by Lo and Chau
\cite{lc97}, some quantum bit commitment protocols were recently
proposed \cite{kent97a,kent97} together with a claim of security that
is ruled out by the general result.  Fortunately, these claims
\cite{kent97} were not published.  In fact, the protocols used the
same idea previously described in \cite{bc96,claude}.
A brief history of the result with proper references to original work
seems appropriate.  We will not describe the proofs again (except in
the Appendix which contains the original proof of Mayers).  Our
objective is to create a wholeness for the different papers written on
the subject.  We will also discuss the general theorem in the context
of the specific ideas and schemes \cite{bc96} which researchers have
tried to realize quantum bit commitment despite this general theorem.

Before we proceed, let us briefly explain the notion of bit commitment
and its impact in quantum cryptography.  Quantum cryptography is often
associated with a cryptographic application called key
distribution~\cite{bb84,bbbss92} and it has achieved success in this
area~\cite{bc96}.  However, other applications of quantum mechanics to
cryptography have also been considered and bit commitment was at the
basis of most if not all of these other
applications~\cite{bc96,bbcs91,bcjl93,yao95}.  A~{\em bit
commitment\/} scheme allows \Sa\ to send something to \Ra\ that
commits her to a bit $b$ of her choice in such a way that \Ra\ cannot
tell what $b$ is, but such that \Sa\ can later prove him what $b$
originally was.  You may think of this as \Sa\ sending a note with the
value $b$ written on it in a strongbox to \Ra\ and later revealing him
the combination to the safe.

\Sa\  can choose the distribution of probability of $b$ during the
commit phase.  The commitment obtained after the commit phase is
binding if \Sa\  cannot change this distribution of probability and it
is concealing if \Ra\  cannot obtain any information about $b$ without
the help of \Sa.  The commitent is secure if it is binding and
concealing.  The commitment is unconditionally secure if it is secure
against a cheater, either \Sa\ or \Ra, with unlimited technology and
computational power. In 1993 a protocol for quantum bit commitment,
henceforth referred to as~BCJL, was claimed to be {\em provably
secure\/}~\cite{bcjl93}, that is, the resulting commitments were
thought to be unconditionally secure.  Because of quantum bit
commitment, the~future of quantum cryptography was very bright, with
new applications such as the identification protocol of Cr\'epeau and
Salvail~\cite{claude:louis} coming up regularly.

The trouble began in October 1995 when Mayers found a subtle flaw in
the BCJL protocol.  Though Mayers explained his discovery to many
researchers interested in quantum bit commitment \cite{BCJL:kaput},
his result was not made entirely public until after Lo and Chau
discovered independently a similar result in March 1996
\cite{lc96}. The result of Mayers was more general than the one
obtained by Lo and Chau, but both used the same basic idea.  The
result of Lo and Chau did not encompass the BCJL protocol in which \Ra\ 
can obtain an exponentially small amount of information. (In practice
a protocol is considered secure as long as \Ra\  cannot obtain more than
an exponentially small amount of information on the bit commited by
\Sa, that is, an amount of information that goes exponentially fast
to $0$ as the number of photons used in the protocol increases.)
However, the final version published by Lo and Chau \cite{lc96} used
the techniques previously used by Mayers \cite{BCJL:kaput} to prove
the non security of the BCJL protocol and any other protocol published
at the time.  So, the paper of Lo and Chau \cite{lc96} is a proper
account of these preliminary results.

\section{The general impossibility theorem}
Now, we review the general theorem \cite{mayers97} (see also the
Appendix) which says that a quantum protocol which creates an
unconditionally secure bit commitment is simply impossible.  The main
additionnal difficulty in the general result is that it is easy to
think that measurements and classical communication could be used to
restrict the behavior of the cheater during the commit phase, and thus
obtain a secure bit commitment.  In fact, after BCJL was shown not
secure, the spontaneous attitude was to try alternative quantum bit
commitment protocols by making some clever use of measurements and
classical communication~\cite{people}.  Some of these protocols were
proposed after that Mayers obtained the general result in March 1996
(just a little bit after Lo and Chau discovered their restricted
result independently).  All of these protocols were found not secure
by Mayers.

There exists two approaches to deal with measurements and classical
communication in quantum bit commitment protocols: an indirect
approach and a direct approach.  In the first proof written by Mayers
(see the Appendix) the indirect approach was used.  It was shown that
any protocol in which classical information is used is equivalent to
another protocol in which no classical information at all is used.
Then it was shown that no protocol of the latter kind is
unconditionally secure.  The first step of this indirect proof is
called a reduction in computer science.  The advantage of this
approach is that, after the reduction is shown, the attack on the new
protocol is easy to describe and analyse because there is no classical
communication anymore.  The disadvantage is that we don't deal
directly with the issue of classical communication and measurements,
that is, the attack obtained against the new protocol is not the one
that applies on the original protocol.  The attack in the new protocol
does not include any classical communication, whereas in the original
protocol the cheater must communicate classically with the honest
participant (otherwise this honest participant will wonder what is
going on).

We emphasize that the proof of the reduction which is not that hard
must nevertheless explain why the cheater can still cheat in the
original protocol despite the fact that he is restricted by
measurements and decoherence which must occur because of classical
communication.  Otherwise the overall proof would simply miss the
important issue of classical communication -- it would not encompass
the protocols and ideas that have been proposed recently
\cite{bc96,claude,kent97a,kent97}.  Because this issue was somehow
confusing, Mayers prefered to use a more direct approach without
reduction in \cite{mayers97}.  So, the paper \cite{mayers97} directly
describes and analyses the real attack that must be executed by the
cheater.

Lo and Chau also wrote a paper \cite{lc97} to discuss the issue of
quantum communication and other aspects of Mayers's result. They used
a variant of Yao's model for quantum communication.  The essence of
Yao's model is that a third system is passed back and fourth under the
control of each participant at their turn~\cite{yao95}.  Mayers's
attack works fine in this model, and it is indeed important to verify
that the attack works in such a reasonable model.  With regard to
classical communication, the discussion of Lo and Chau \cite{lc97} is
similar to the indirect approach.

Now, let us consider the attack.  Of course, we are interested in the
attack on the original protocol.  The attack on the new protocol is
just a construction in a proof.  We emphasize that in both approaches,
with a reduction or without a reduction, the attack on the original
protocol is the same.  Here we focus on the part of the attack which
must be executed during the commit phase.  (The remainder of the
attack which is executed after the commit phase is the same as when
there is no classical communication, so it creates no additional
difficulty.)  One ingredient in the attack is that the cheater keeps
every thing at the quantum level except what must be announced
classically.  Assume that at some given stage of the commit phase, a
participant has normally generated a classical random variable $R$,
executed measurements to obtain an overall outcome $X$, and shared
some classical information $Y$ with the other participant as a result
of previous communication.  Now, assume that this participant is the
cheater and that the protocol says he must transmit some classical
information $f(X,R,Y)$, which for simplicity we assume is a binary
string.  One might think that the cheater must have generated $X$ and
executed the measurements, or at the least some of them, in order to
be able to compute and send $f(X,R,Y)$.  However, the cheater does not
have to do that.  He can do the entire computation of $f(X,R,Y)$,
including the computation of $X$ and the measurements, at the quantum
level.  Only $Y$ needs to be classical.  Then he can measure the bits
of the string $f(X,R,Y)$ (only these bits) and send them to the other
participant. An example is given in section \ref{an_example_section}.
The final result is that every information is kept at the quantum
level, except what must be sent classically to the other
participant. As explained in \cite{dom:impos,mayers97} (see also the
Appendix) this strategy executed during the commit phase either allows
\Ra\  to obtain some information about the bit commited by \Sa,
without any help from \Sa, or else allows \Sa\  to change her mind
after the commit phase (as in the example of section
\ref{an_example_section}).

This is not the end of the story.  After that the above argument was
understood, Cr\'epeau proposed a quantum protocol \cite{bc96,claude}
that uses a computationally secure classical bit commitment
\cite{BCC88,NOVY} as a subprotocol.  The idea was to rely temporarily
on the limitation (in speed) on the cheater during the commit phase to
force him to execute some measurements (or restrict his behavior in
some other way) in order to obtain a secure bit commitment.  The hope
was that this short-term assumption could be dropped after the commit
phase so as to obtain a quantum bit commitment not relying on any
long-term assumption.  The same idea was recently used by Kent in
\cite{kent97}.  Salvail also proposed a protocol in which two
participants, \Sa\ and \Saa\ say, want to commit a bit to \Ra.  \Sa\
and \Saa\ are sufficiently far apart that they cannot communicate
during the commit phase.  Again the hope was that this temporary
restriction on the cheaters during the commit phase would be
sufficient to obtain a secure quantum bit commitment not relying on
any long-term assumption.

However, after some thoughts, one realize that the cheater in Mayers's
attack executes the honest algorithm, the only difference is that he
executes this honest algorithm at the quantum level.  Therefore, if
the cheater has the power to execute the honest protocol (which he
must have) and has the technology to store information at the quantum
level, then he has the power to cheat during the commit phase, despite
the fact that he has not the power to break the computationally secure
bit commitment efficiently, or despite the fact that \Sa\  and
\Saa\  cannot communicate during the commit phase.  After the
commit phase, the rule of the game is that we must drop the assumption
on the computational power of the cheater, so the fact that a
computationally secure bit commitment was used is irrelevant: the
proof applies.


\section{An Example:How to Break Kent's Protocol}
\label{an_example_section}
In this section we illustrate the discussion of the previous section
by a concrete example.  We shall show how to break Kent's proposal
\cite{kent97a,kent97} for a quantum bit commitment scheme using a
time-bounded computational assumption.  The paper \cite{kent97}
describes two constructions for such a scheme, one allows \Sa\  to
commit and the other allows \Ra\  to commit permanently. In this section
we break the protocol allowing \Sa\  to commit permanently. The other
version can be broken by a similar attack.

Kent's protocol \cite{kent97} uses a classical and unconditionally
hidding bit commitment scheme. The hope is that this classical scheme
will constraint \Sa\  to transmit q-bits in pure states.  The protocol
uses the BB84 coding scheme: $\Psi(0,0) = \ket{0}_{+}, \Psi(0,1) =
\ket{1}_{+}, \Psi(1,0) = \ket{0}_{\times}, \Psi(1,1) =
\ket{1}_{\times}$.  The first bit corresponds to the basis and the
second bits to the encoded bit. Here is the essential idea behind
Kent's protocol. For each $i = 1, \ldots, N_B$, \Sa\  picks a random
pair $(x,z) = (x_i,z_i) \in \{0,1\}^2$, sends a photon $\pi_i$ in the
BB84 state $\Psi(x,z)$ and execute a classical bit commitment
$BC(x,z)$ according to the above classical bit commitment scheme.
We denote\footnote{Notation $\{a,b\}_{[s]}$ for $s\in\{0,1\}$ is $a$
if $s=0$ and $b$ if $s=1$.}  $\theta_i = \{+,\times\}_{[x]}$ the basis
used by \Sa\  for the photon $\pi_i$.  \Ra\  then picks a random sample
$X \subseteq \{1,\ldots,N_B\}$ of size $N_B - N$.  For each $i\in X$,
\Ra\  asks \Sa\  to unveil $(x,z)\in\{0,1\}^2$ corresponding to the
committed pair of classical bits in $BC(x,z)$. \Ra\  then measures
$\pi_i$ in basis $\theta_i = \{+,\times\}_{[x]}$ and verifies that the
observed outcome is indeed $z\in\{0,1\}$.  The idea behind the
remainder of Kent's protocol is very similar to the first bit
commitment scheme ever proposed by \cite{bb84}.  The difference is
that in \cite{bb84} \Sa\  picks the same value for all $x_i$, that is,
the string of bases used by \Sa\  is either $+ + \ldots +$ or $\times
\times \ldots \times$. (See also \cite{mayers97,lc96} for a
description and analysis of this protocol.)  The basic idea is that
the committed bit is encoded in the transmission basis for each photon
$\pi_i$. In Kent's protocol, if \Sa\  wants to commit bit $b$, she
announces $x_i \oplus b$ for each $i \in Y = \{1,\ldots,N_B\} - X$.
So, the bit is commited in the choice of basis used by \Sa\  for each
$i \in Y$.  

The scheme is unconditionnally hidding because no information about
the transmission basis can be obtained from any photon $\pi_i$ since
the density matrix corresponding to the transmission in rectilinear
basis
\[ \rho_{+}=\frac{1}{2}\ket{0}_{+}\bra{0}_{+}+\frac{1}{2}\ket{1}_{+}\bra{1}_{+}
\]
and the one corresponding to the transmission in diagonal basis
\[ \rho_{\times}=\frac{1}{2}\ket{0}_{\times}\bra{0}_{\times}+
\frac{1}{2}\ket{1}_{\times}\bra{1}_{\times}
\]
are such that $\rho_{+}=\rho_{\times}$. 

Clearly, if sends the pure states $\Psi(x_i,z_i)$, she cannot claim
that she used the other basis, that is, the one associated with $x_i
\oplus 1$, for each $i \in Y$. So, if really \Sa\ has sent the pure
states $\Psi(x_i,z_i)$, the protocol should be binding.  \Sa\ can
cheat in the original protocol of \cite{bb84} by sending EPR pairs
rather than a mixture of BB84 quantum states (see \cite{bb84} for more
details).  So the resulting commitment is not binding.  In Kent's
protocol, if \Sa\ cannot break the computational assumption during
this test phase (between the time the commitments have been sent and
the time they are opened), it is argued that \Ra\ gets convinced that
almost all photons $\pi_i$ in $Y$ are in the pure states
$\Psi(x_i,z_i)$.  Indeed, if this was true then the protocol would
also be binding.  However, we show that it is not the case.

\subsection{The Classical Commitment}
Let us first model the classical and unconditionally hiding
commitment scheme by four one-way permutations
\footnote{The same kind of argument can also be formalized for general
one-way functions rather than one-way permutations.  However, no
classical and unconditionally bidding bit commitment scheme is yet
known to be based only on the existence of one-way functions.}
$f_{00},f_{01}, f_{10}, f_{11}:\{0,1\}^n\rightarrow \{0,1\}^n$ for any
integer $n$. In the remaining, functions $f_{00},f_{01},f_{10}$ and
$f_{11}$ need not to be distinct.
To commit $(x,z)$, \Sa\  picks a random uniformly distributed $w \in
\{0,1\}^n$ and sends $y = f_{xz}(w)$ to \Ra.  We obtain that $y$, the
piece of evidence that \Sa\  gives to \Ra\  in order to commit on a pair
of classical bit $(x,z)$, is a random element uniformly distributed
in $\{0,1\}^n$.  Here are the properties that we need to consider.
\begin{enumerate}
\item The functions $f_{xz}$ are efficiently computable and
      publicly known. 
\item Given $y = f_{xz}(w)$ no information on $(x,z)$
      is known by \Ra\  (thus the protocol is unconditionally hiding).
\item \Sa\  knows only one $(x,z,w)\in\{0,1\}^2 \times \{0,1\}^n$ such
      that $f_{xz}(w)=y$. If she manages to find another $(x',z',w')
      \in\{0,1\}^2 \times \{0,1\}^n$ distinct from $(x,z,w)$ such that
      $f_{x'z'}(w') = y$ then she can break the computational
      assumption (because necessarily $(x,z) \neq (x',z')$).
\end{enumerate}
We shall see that the above conditions for classical commitment, in
particular the first two conditions, implies that the proposed method
cannot ensure \Ra\  that most of the remaining q-bits are in pure
states.  We have described a particular classical bit commitment
scheme, but Mayers's attack works with any other classical bit
commitment scheme.  In the next two subsections we describe \Sa's
attacks during the commit phase, then in the third subsection we
explain how \Sa\  can change her mind after the commit phase.

\subsection{Alice's Preparation}
If \Sa\  wants to cheat the proposed protocol, as we will see, she has
only to send entangled states rather than a mixture of BB84 states.
In Kent's protocol, the use of a classical bit commitment scheme is
intended to rule out the EPR attack. However, other entanglements can
do the job.  Let us consider the state $\ket{\gamma(\theta)}$ defined
upon \footnote{In the following we sometime consider $\theta\in\{+,\times\}$
as being the bit $x$ such that $\theta=\{+,\times\}_{[x]}$. Notations
$f_{+ z}(x)$ means $f_{0z}(x)$ and $f_{\times z}(x)$ means $f_{1z}(x)$.}
 $\theta \in\{+,\times\}$ as \footnote{When a quantum state is
written as $\ket{w}$ for $w\in\{0,1\}^n$ we mean $\ket{w_1}_+ \otimes
\ldots \otimes \ket{w_n}_+$.}
\begin{equation}\label{eq}
\ket{\gamma(\theta)}= \frac{1}{\sqrt{2^{n+1}}}
\sum_{w\in\{0,1\}^n}
\ket{w}\ket{f_{\theta 0}(w)}\ket{0}_{\theta}\ket{{0}}_{\theta}
+
\ket{w}\ket{f_{\theta 1}(w)}\ket{1}_{\theta}\ket{{1}}_{\theta}.
\end{equation} 
Mayers's theorem also specifies that there should be a superposition
over $x$ (or equivalently over $\theta$).  The idea is that every
random choice, including the choice of the bases, must be done at the
quantum level.  However, this part of the superposition would collapse
immediately because \Sa\  must announce $x_i \oplus b$, for the
classical bit $b$ she has chosen (this is what is specified by the
attack).  So for simplicity we ignore this part of the superposition.

The state (\ref{eq}) can be efficiently constructed from condition 1
about the classical commitment scheme.  The state
$\ket{\gamma(\theta)}$ is made out of four registers which we denote
from left to right as $r_w,r_f,r^A_z$ and $r^B_z$. Now suppose \Sa\ 
sends the register $r^B_z$ to \Ra\  instead of a random BB84 pure
state. We assume the more general case where \Ra\  does not measure the
received quantum states until the pure states verifications take
place. This allows a more reliable test than measuring immediately
after reception and testing later on. Let $H_A$ be the Hilbert space
for registers $r_w,r_f$ and $r^A_z$ and let $H_B$ be the Hilbert space
for register $r^B_z$.  By construction, \Ra\  receives a mixture with
density matrix:
\begin{equation}\label{pure}
\rho_B=\trp{H_A}{\ket{\gamma(\theta)}\bra{\gamma(\theta)}}=\rho_{\theta}.
\end{equation}
\Sa's preparation consists of $N_B$ systems $s_1,\ldots,s_{N_B}$ in
quantum states
$\ket{\gamma(\theta_1)},\ldots,\ket{\gamma(\theta_{N_B})}$ for
$\theta_i \in_R \{+.\times\}$.  She sends to \Ra\  the $r^B_z$ registers
for all systems $s_1,\ldots,s_{N_B}$.

\subsection{How Alice  Deals With Classical Communication}
Suppose \Sa\  has sent all $N_B$ registers $r^B_z$ to \Ra. Let
$\theta_1,\ldots,\theta_{N_B}$ be the $N_B$ bases picked in
$\{+,\times\}$ in order to prepare the states
$\ket{\gamma(\theta_1)},\ldots,\ket{\gamma(\theta_N)}$. (From \Sa's
point of view these bases, i.e. the $x_i$, are not random anymore.)
To execute the classical commitment, \Sa\  must send the classical
information $f_{xz}(w)$.  Thus far, the values of $z$ and $w$ are not
fixed: they are still in superposition.  As explained in the previous
section, \Sa\  does not have to obtain $w$ nor $z$ classically to
compute $f_{xz}(w)$.  For committing, \Sa\  simply measures in
rectilinear basis all registers $r_f$.  She then announces to \Ra, for
each $i\in\{1,\ldots,N\}$, the result $y_i$. That is the way Mayers's
attack works.  Each system $s_i$ is now in state
\[ 
\ket{\gamma'(\theta_i)}=\frac{1}{\sqrt{2}}\left(
\ket{w}\ket{y_i}\ket{0}_{\theta_i}\ket{0}_{\theta_i}+
\ket{w'}\ket{y_i}\ket{1}_{\theta_i}\ket{1}_{\theta_i}
\right)
\]    
where $w = f^{-1}_{\theta_i 0}(y_i)$ and $w' = f^{-1}_{\theta_i
1}(y_i)$.  The above state is guaranteed to occur by property 2 of the
classical commitment scheme, and the fact that $w$ is uniquely
determined by $x$ (or $\theta$), $z$ and $y$.

Now suppose \Ra\  asks \Sa\  to unveil the commitment for some position
$i\in X$. \Sa\  simply measures registers $r_w$ (in basis $+$) and
$r^A_z$ (in basis $\theta_i$) for the system $s_i$.  Let $w$ and $z$
be the outcomes of the measurement.  \Sa\  announces $w$, $z$ and
$x_i$ to \Ra. \Ra\  always verifies that $y_i=f_{x_iz}(w)$. The system
ends up in state
\[
\ket{\gamma''(\theta_i)}=\ket{w}\ket{y_i}\ket{z}_{\theta_i}\ket{z}_{\theta_i}
\]
which leads to a successful verification by \Ra. Clearly \Sa\ can
always pass the test without breaking the computational assumption.
The main point is that \Sa\ executes the honest protocol at the
quantum level, so any computational bound is useless.  It follows that
Kent's verification procedure is not a verification that almost all
received q-bits are in pure states since equation \ref{pure} is
obviously not the description of a pure state.

\subsection{Breaking the Quantum Scheme}
We now show how \Sa\  can decide freely the bit she wants to unveil.
We recall that, at this point, the rule of the game is that all
computational assumptions must be dropped. (Otherwise we only have a
computationally secure bit commitment, and this can already be done
classically.)  After the verification procedure only the remaining
systems $s_i$ with $i\in Y=\{1,\ldots,N\}\setminus X$ are used. In
order to break the quantum protocol it is sufficient to show how \Sa\ 
can choose the transmission basis for all photons transmitted to
\Ra. For all $i\in Y$ the system $s_i$ is in state (we remove the
$r_f$ register since it is no more entangled but we remember its
observed value $y_i$):
\[
\ket{\gamma'(\theta_i)}=\frac{1}{\sqrt{2}}\left(
\ket{w}\ket{0}_{\theta_i}\ket{0}_{\theta_i}+
\ket{w'}\ket{1}_{\theta_i}\ket{1}_{\theta_i}\right).
\]
To cheat, \Sa\  must disentangle the register $r_w$ and obtain
the state
\[
\ket{\gamma''(\theta_i)}=\frac{1}{\sqrt{2}}\left(
\ket{w_0}\ket{0}_{\theta_i}\ket{0}_{\theta_i} +
\ket{w_0}\ket{1}_{\theta_i}\ket{1}_{\theta_i}\right).
\]
where $w_0$ is some fixed string.  If we ignore the disentangled
registers $r_f$ and $r_w$, this state is essentially an EPR pair
(modulo a unitary transformation on \Sa's side).  So \Sa\ can cheat as
in the original attack defined in \cite{bb84}.  Now, we show how \Sa\
can disentangle $r_w$.  A simple way to disentangle $r_w$ would be to
replace both $w$ and $w'$ by the same output $f_{\theta_i 0}(w) =
f_{\theta_i 1}(w') = y_i$.  This is a reversible computation executed
in the computational basis defined by $\theta_i$ for $r^A_z$ and $+
\ldots +$ for $r_w$, so it corresponds to a unitary transformation.
This answers the question.  However, this answer is somehow misleading
because it gives the impression that the attack is as simple as the
computation of $f_{\theta_i 0}(w) = f_{\theta_i 1}(w') = y_i$.  The
problem is that one must still explain how the two distinct inputs $w$
and $w'$ can be replaced by one and the same value $y_i$. Here we show
how this can be done by \Sa\ if she can inverse the functions
$f_{\theta_i z}$.  Because she knows $y_i$ and $\theta_i$, she can
compute $w = f^{-1}_{\theta_i 0}(y_i)$ associated with
$\ket{0}_{\theta_i}$ and $w' = f^{-1}_{\theta_i 1}(y_i)$ associated
with $\ket{1}_{\theta_i}$, so she can ``erase'' the register $r_w$,
that is, she can set this register to ${\bf 0}$ by a bit-wise addition
modulo 2.  (She can also set it to any other value she wants,
including $y_i$.)  This concludes the description of the attack.

Although \Sa\ breaks the computational assumption (i.e. inverse
the functions $f_{x z}$) in order to unveil the bit she wishes, this
cannot be used as a building block for a secure quantum bit commitment
where the computational assumption is no more needed after some
time. This is for exactly the same reason than the one allowing to
conclude that no quantum bit commitment can be built from a classical
computational assumption.


\section{Conclusions}
The first proof provided for the impossibility of bit commitment (see
the Appendix) has completely obliterated the possibility of creating an
unconditionally secure bit commitment.  However, the attack was only
indirectly described.  Subsequently, specific attempts to by-pass this
general result were proposed\cite{bc96,claude}.  This has shed more
light on the nature of the attack which was finally described
explicitly in \cite{mayers97}.  Our goal here was to provide an
analysis of this general attack in the context of a specific example,
and to create a wholeness for the different papers published on the
subject. The big lesson to learn from all this is that quantum
information is always more elusive than its classical counterpart:
extra care must be taken when reasoning about quantum cryptographic
protocols and analyzing them.  We hope that this paper will help to
clarify the issue of the impossibility of bit commitment in its full
generality.

\newpage
\section*{Appendix}
This appendix contains the original proof written by Mayers and sent
to few researchers by email on March 14 1996.  A modified version of
the proof, which also used a reduction, was published in
\cite{dom:impos}.  A direct proof with no reduction was published
later in \cite{mayers97}.

\begin{abstract}
It is currently known that the 1993 BCJL protocol 
of Brassard, Cr\'epeau, Jozsa and Langlois (BCJL) is insecure.
Here we provide the first proof that, not only this protocol, 
but any quantum bit commitment is either insecure
against \Sa\  or insecure against \Ra.  
\end{abstract}

\setcounter{section}{0}

\section{Introduction}     
The fact that the quantum bit commitment protocol 
of Brassard, Cr\'epeau,  Jozsa and Langlois \cite{bcjl93} is
insecure is known for quite sometime~\cite{BCJL:kaput}.
Lo and Chau have also independently shown that a restricted 
category of quantum bit commitments is insecure~\cite{lc96}.
Now, we provide the first 
proof that not only these quantum bit commitment protocols,
but any other quantum bit commitment protocol
is insecure.  

The absence of quantum bit commitment is 
a serious concern because other quantum protocols such
as quantum oblivious transfer depend on the security of bit 
commitment~\cite{bbcs91,crepeau94,cgt95,yao95}.  
On~the other hand, not all of Quantum Cryptography fall apart
because our earlier proof of security for quantum
key distribution~\cite{mayers95a}  holds even if secure quantum
bit commitment is not possible despite the fact that it is
based on an earlier ``proof'' of security for quantum oblivious transfer
that fails in the absence of a secure bit commitment scheme.
The reason is that the proof of security for quantum key distribution
does not depend on the security of quantum oblivious transfer, but
rather on the (correct) proof that quantum oblivious transfer would
be secure if implemented on top of a secure bit commitment scheme.
     
\section{Bit Commitment}     
Any cryptographic task defines the relationship between      
inputs and outputs respectively     
entered and received by the task's participants.       
In bit commitment, \Sa\  enters a bit $b$.     
At a later time,  \Ra\  may request this bit and,     
whenever he does, he receives this bit,      
otherwise he learns nothing about $b$.      
     
In a naive but concrete realization of bit commitment,     
\Sa\  puts the bit into a strong-box     
of which she keeps the key and then     
gives this strong-box to \Ra.  At a later time,     
if \Ra\  requests the bit, \Sa\  gives the key to \Ra.       
The main point is that \Sa\  cannot change her mind     
about the bit $b$ and \Ra\  learns nothing about it unless he     
obtains the key.

\section{Quantum Bit Commitment: the attack}     
For every quantum bit commitment protocol $Q$, we
shall construct a protocol $\widetilde{Q}$, show that
the security of $Q$ implies the security
of $\widetilde{Q}$ and then show that $\widetilde{Q}$ is insecure.  

Let $A$ and $B$ stand for \Sa\  and \Ra\  respectively.
For any bit commitment protocol $Q$, the state space $H$ 
is of the form $H_A \otimes H_B$ where
$H_A$ and $H_B$ are state spaces 
on \Sa's side and \Ra's side respectively.  
\Sa's and \Ra's generation of classical
variables, measurements, unitary transformations, etc in the
commit phase of $Q$ can be modeled by two global measurements, one
on \Sa's side and the other one on \Ra's side.  These two measurements
together correspond to an overall measurement on the entire state 
space $H = H_A \otimes H_B$. This single overall measurement
corresponds to the entire commit phase of $Q$.  

Now, we construct $\widetilde{Q}$.  For every $P \in \{A,B\}$, the
state space on $P$'s side is of the form $\widetilde{H}_P = H_P
\otimes H'_P$.  The entire state space is $\widetilde{H} =
\widetilde{H}_A \otimes \widetilde{H}_B$.  The additional parts $H'_A$
and $H'_B$ are used to store the outcome of the overall measurement
executed by \Sa\  and \Ra\  together, that is, the overall measurement
executed by \Sa\  and \Ra\  in the commit phase of $Q$ becomes a unitary
transformation on $\widetilde{H}$.  At the opening phase (or just
after the commit phase), \Ra\  and \Sa\  obtain the classical variables
stored in their respective systems $H'_A$ and $H'_B$, that is they
execute the measurements that they normally execute in $Q$, and they
continue with the opening phase as in $Q$.

It is not hard to see that the non security of $\widetilde{Q}$ implies the 
non security of $Q$.  Assume that $P \in \{A,B\}$ can cheat in $\widetilde{Q}$.
A dishonest $P$ in $Q$ can do exactly as $P$ in $\widetilde{Q}$.     
The resulting random situation in $Q$  after 
the commit phase is the same random situation that holds in $\widetilde{Q}$ 
after that the other participant $\bar{P}$ has measured
his quantum system $H'_{\bar{P}}$. 
So, if $P$ succeed in $\widetilde{Q}$, $P$ also succeed in $Q$.  

Now, we must show that $\widetilde{Q}$ is insecure.  
It is a principle that we must assume that every 
participant knows every detail of the protocol, including
the distribution of probability of a random variable generated
by another participant.
There is no loss of generality in assuming that at the beginning
of the protocol, the overall system 
is in a pure state $|\psi\rangle \in \widetilde{H} 
= \widetilde{H}_A \otimes \widetilde{H}_B$: 
the preparation of a mixture could be 
included as a part of the protocol.
The commit phase of the protocol specifies a unitary 
transformation $U_b$ on the entire system.  So at the
end of the commit phase, the overall
system is in a final state $|\phi_b\rangle = U_b |\psi\rangle$.

It is fair to assume that every thing outside $\widetilde{H}_A$ 
is under the control of a dishonest \Ra.  In other words,
there are no third system $\widetilde{H}_C$. 
For $b =0,1$, let $\rho^A_b$ and $\rho^B_b$
be the partial traces of $|\phi_b\rangle\langle \phi_b|$
over $\widetilde{H}_B$ and $\widetilde{H}_A$ respectively.
The density matrices
$\rho^B_0$ and $\rho^B_1$ on \Ra's side must be 
close one to the other, otherwise \Ra\  can cheat.  
We shall do the simpler case $\rho^B_0 = \rho^B_1$.
The more subtle case where the density matrices
are not identical is done in the next section.

Consider the Schmidt decomposition \cite{hughston93,schmidt}
of $|\phi_0\rangle$ and $|\phi_1\rangle$ 
respectively given by
\[
|\phi_0\rangle = \sum_i \sqrt{\lambda_i} |e^{(0)}_i\rangle
\otimes |f_i\rangle
\]
and
\[
|\phi_1\rangle = \sum_i \sqrt{\lambda_i} |e^{(1)}_i\rangle
\otimes |f_i\rangle
\]
In the above formula, 
$\lambda_i$ are eigenvalues of the density matrices
$\rho^B$, $\rho^A_0$ and $\rho^A_1$.
The fact that these density matrices share the same positive
eigenvalues with the same multiplicity is part of 
the Schmidt decomposition theorem \cite{hughston93,schmidt}.  
The states $|e^{(b)}_i\rangle$ and $|f_i\rangle$
are respectively  eigenstates of $\rho^A_b$ and $\rho^B$
associated with the same eigenvalue $\lambda_i$.
Clearly, the same unitary transformation that maps
$|e^{(0)}_i\rangle$ into $|e^{(1)}_i\rangle$
also maps $|\phi^{(0)}\rangle$ into $|\phi^{(1)}\rangle$.
We recall that \Sa\  knows what are
the states $|\phi_0\rangle$ and $|\phi_1\rangle$. Therefore,
she can determine the above unitary transformation.

In order to cheat, \Sa\ creates the state $|\phi_0\rangle$.  In other
words, \Sa\ does what she must honestly do when she has $b=0$ in mind.
With the state $|\phi^{(0)}\rangle$ \Sa\ is able to convince \Ra\ that
she had $b = 0$ in mind:
\Sa\  has only to open the bit as an honest \Sa\  
would in $\widetilde{Q}$ with $b=0$ in mind.  If \Sa\ want to change
her mind, she only has to maps $|\phi^{(0)}\rangle$ into
$|\phi^{(1)}\rangle$ before she continue the opening phase as if she
had $b = 1$ in mind.
      
\section{The real situation}     \label{realcase}
Now, we consider the real situation
where the density matrices $\rho_0$ and $\rho_1$ are      
not identical. If the protocol is to be secure against \Ra, the     
density matrices $\rho_0$ and $\rho_1$      
must respect some constraint.       
We express this constraint in terms of      
measurements on the $n$ photons that     
return a binary classical outcome $X \in \{0,1\}$.     
We recall that \Sa\      
prepares the density matrix $\rho_b$ when she has     
$b$ in mind, that is, when $B = b$.     
Without loss of generality,      
we take the convention  that
\mbox{$\Pr{X = 0 | B = 0} \geq \Pr{X = 0 | B = 1}$}.     
We~denote $X_b$ the random variable $X$ conditioned by $B = b$     
so that \mbox{$\Pr{X = x | B = b} = \Pr{X_b = x}$}.       
The constraint is      
\[     
\left| \; \frac{1}{2} -  PE \; \right| =      
\left|\; \frac{1}{2}  - \sum_{b = 0}^1 \Pr{B = b} \Pr{X_b = \bar{b}} \;\right| 
\leq 2^{-\alpha n}.     
\]     
This constraint says that no matter which measurement \Ra\ uses to
decide between $B = 0$ and $B = 1$, the probability of error is
exponentially close to $1/2$.  It has been shown in
\cite{BCJL:kaput,mayers96thesis}, building on the work of
\cite{hughston93,josza94}, that this implies the existence of two
purifications $|\psi_0\rangle$ and $|\psi_1\rangle$ for $\rho^B_0$ and
$\rho^B_1$ respectively such that
\[     
\langle \psi_0 | \psi_1 \rangle^2  \geq (1 - 2 \times 2^{-\alpha n}). 
\]     
We have that $|\psi_0\rangle$ and $|\psi_1\rangle$ 
are almost the same state.  

In order to cheat, \Sa\  prepares the state $|\psi_0\rangle$.
If she want to unveil $b = 0$, using the same
argument as in the simpler case, 
she maps $|\psi_0\rangle$ into $|\phi_0\rangle$
and continue as in the honest $\widetilde{Q}$ 
when she has $b =0$ in mind.  If she
wants to unveil $b =1$, she executes on $|\psi_0\rangle$ 
the unitary transformation $F$ that would map
$|\psi_1\rangle$ into $|\phi_1\rangle$. 
She obtains the state $F |\psi_0\rangle$.  The inner
product between the desired state 
$|\phi_1\rangle = F|\psi_1\rangle$
and the actual state $F|\psi_0\rangle$, is the same as the
inner product $\langle \psi_1 | \psi_0\rangle$
which is exponentially close to $1$.  So, for all
practical purpose, \Sa\  can cheat as in the simpler
case by applying this transformation $F$ and then
continuing as in the honest $Q'$ when she has $b =1$ in mind.
This concludes the proof that every bit quantum bit commitment
is insecure.
     
Note that as~a consequence, Yao's proof of security for Quantum     
Oblivious Transfer~\cite{yao95} fails because it is built on insecure
foundations (through no fault of Yao).  Ironically, as we stated in the
Introduction, the proof of security for Quantum Key Distribution shown in
\cite{mayers95a} stands despite the fact that it draws on Yao's work
because it does not depend on the security of Bit Commitment.

\end{document}